\documentstyle[aps, draft, epsf, floats]{revtex}

\preprint{UCRHEP-T193}

\newcounter{opdnum}
\newcounter{opfnum}

\def\opd{\addtocounter{opdnum}{1} O_D^{(\theopdnum)}}
\def\opf{\addtocounter{opfnum}{1} O_F^{(\theopfnum)}}

\newcommand{\U}{U^c}
\newcommand{\D}{D^c}
\newcommand{\E}{E^c}
\newcommand{\Ho}{H_1}
\newcommand{\G}{H_2}

\newcommand{\Qd}{Q^\dagger}
\newcommand{\Qe}{(Q^\dagger e^{V_Q})}

\newcommand{\Ud}{U^{c\dagger}}
\newcommand{\Ue}{(U^{c\dagger} e^{V_U})}

\newcommand{\Dd}{D^{c\dagger}}
\newcommand{\De}{(D^{c\dagger} e^{V_D})}

\newcommand{\Ld}{L^\dagger}
\newcommand{\Le}{(L^\dagger e^{V_L})}

\newcommand{\Ed}{E^{c\dagger}}
\newcommand{\Ee}{(E^{c\dagger} e^{V_E})}

\newcommand{\Hd}{H_1^\dagger}
\newcommand{\He}{(H_1^\dagger e^{V_{H_1}})}

\newcommand{\Gd}{H_2^\dagger}
\newcommand{\Ge}{(H_2^\dagger e^{V_{H_2}})}

\newcommand{\qt}{\tilde{q}}

\newcommand{\ub}{u^c}
\newcommand{\ut}{\tilde{u^c}}

\newcommand{\db}{d^c}
\newcommand{\dt}{\tilde{d^c}}

\newcommand{\lt}{\tilde{l}}

\newcommand{\eb}{e^c}
\newcommand{\et}{\tilde{e^c}}

\newcommand{\ho}{h_1}
\newcommand{\hot}{\tilde{h}_1}

\newcommand{\gt}{h_2}
\newcommand{\gtt}{\tilde{h_2}}

\newcommand{\lG}{\lambda_G}
\newcommand{\lW}{\lambda_W}
\newcommand{\lB}{\lambda_B}

\newcommand{\p}{\phi}
\newcommand{\th}{\theta}

\newcommand{\tstbs}{{\theta^2\bar{\theta}^2}}
\newcommand{\al}{\alpha}

\newcommand{\da}{\dagger}

\newcommand{\e}{\epsilon}

\newcommand{\ns}{\nabla^2}
\newcommand{\nsb}{\bar{\nabla}^2}
\newcommand{\na}{\nabla^\alpha}
\newcommand{\nas}{\nabla_\alpha}

\newcommand{\La}{{\cal{L}}}

\newcommand{\pa}{\partial}
\newcommand{\bD}{\bar{D}}

\newcommand{\gev}{\hbox{ GeV}}
\newcommand{\tev}{\hbox{ TeV}}
\newcommand{\lesim}{\stackrel{\sim}{<}}
\newcommand{\gesim}{\stackrel{\sim}{>}}

\begin{document}

\title{Effective operators in Supersymmetry}
\author{Dardo P\'{\i}riz and Jos\'{e} Wudka }
\address{Department of Physics \\
University of California-Riverside \\
California, 92521-0413\\
\vspace{.3cm}
%\rm{PRELIMINARY VERSION, UCRHEP-T179}
}
\date{\today}

\maketitle

\begin{abstract}

We consider a low-energy supersymmetric scenario in which the effects
of heavy supersymmetric interactions are included in a
model-independent manner through a series of supersymmetric
$SU(3)_C\times SU(2)_L\times U(1)_Y$ invariant operators; we explicitly
construct all such operators of dimension $ \le 6 $. From this set, those
operators generated at tree level by the underlying theory are isolated
since under weak coupling conditions they are expected to provide the
largest effects in low-energy processes.
Potential deviations from low energy predictions in process such as proton
decay,
neutrinoless double-$\beta $ decay and flavor changing neutral currents are
analyzed as illustrations of the
method.

\end{abstract}

\bigskip\bigskip
%\newpage
%%%%%%%%%%%%%%%%%%%%%%%%%%%%%%%%%%%%%%%%%%%%%%%%%%%%%%%%%%%%%%%%%%%%%%%%%%%%%%

\section{Introduction}

%%%%%%%%%%%%%%%%%%%%%%%%%%%%%%%%%%%%%%%%%%%%%%%%%%%%%%%%%%%%%%%%%%%%%%%%%%%%%

The Standard Model of particle physics has been successfully tested to
great accuracy in the past decade~\cite{SM},
the latest result being the remarkable
discovery of the top quark~\cite{top} at the Tevatron.
Despite the many
achievements of the Standard Model at current energies,
there are reasons to believe it is but the low-energy manifestation
of a more fundamental theory~\cite{beyo}.
Supersymmetry provides an appealing extension of the Standard Model,
which among other virtues, solves
the gauge hierarchy problem naturally~\cite{hier}. Another
argument in favor of supersymmetry is the predicted unification
of all gauge coupling constants at approximately $10^{16}$
GeV, which can be taken as a hint of supersymmetric grand
unification~\cite{susyco}.

In this paper we study a scenario in which low-energy supersymmetry has
been found and its details have been determined. Since such low-energy
supersymmetric theories often contain indications of some deeper symmetry
(such as the unification of the coupling constants mentioned above), we
will assume that such a supersymmetric theory represents the low-energy
limit of a more fundamental supersymmetric theory (such as
theories exhibiting supersymmetric grand unification~\cite{susyuni},
supersymmetric
left-right models~\cite{leri}, etc.). We will then look for
signatures of this more fundamental theory using an effective Lagrangian
approach~\cite{effop,jose} which will allow us to isolate the
observables in the low-energy supersymmetric theory which are most
sensitive to the virtual effects of the underlying theory. In our
calculations we will not
assume a specific high-energy supersymmetric model; we will instead
determine the kinds of physics probed
by the different observables which we consider and their corresponding scales.
We assume that the heavy theory is weakly coupled.

This procedure has been carried out for some particular interactions, such
as those relevant to proton decay~\cite{wsy}. In this publication we extend
these
results by obtaining the complete list (for a given low-energy
supersymmetric spectrum) of dimension 5 and 6 operators. We will
study the effects of these effective interactions on lepton
and baryon number violation and flavor changing processes.

In order to pursue the above program we need to choose a low energy
supersymmetric theory. We will use, for simplicity, the minimal extension of
the
Standard Model (MSSM)~\cite{mssm}. The whole program can be carried out
for any other low-energy supersymmetric model (though the presence of new
fields
considerably lengthens the computations). In the following we will only
use the particle content of the MSSM, constraints such as the
relationships between the scalar and vector masses~\cite{mssm}
will not be imposed.

In the MSSM each Standard Model particle acquires
a supersymmetric partner of the same mass. Concerning the scalar sector,
(at least) two Higgs doublets
with their corresponding supersymmetric partners are required
due to the holomorphic character of the superpotential
and to achieve anomaly cancelations.
In this theory, baryon number is not naturally conserved, which leads
to an unacceptably fast rate for proton decay. In order to avoid this
problem a new symmetry (R-parity) must be imposed~\cite{wsy,parity}.
After this is done one finds that the most general renormalizable
supersymmetric potential compatible with all the symmetries
does not break gauge and super-symmetries.
This necessitates the introduction of
the so-called ``soft breaking terms''~\cite{soft}, {\it i.e.},
super-renormalizable terms which
break supersymmetry and electroweak symmetry.
The fact that such terms are needed supports the idea that the MSSM is
the low-energy effective model of a more fundamental theory which becomes
manifest above a scale $M$~\cite{sugra}.

The low-energy effective Lagrangian of a weakly coupled renormalizable
theory can be written as a linear combination of operators in the light
fields suppressed by powers of the heavy scale $M$ (as guaranteed by
the decoupling theorem~\cite{dec}). In our case each of these operators
is built from the light superfields and respects the low-energy
symmetries ({\it i.e.} $SU(3)_C\times SU(2)_L\times U(1)_Y
\times $R-parity). The detailed characteristics of the full theory will
be encoded in the set of coefficients multiplying the operators. It is
useful to divide the operators into those that can be generated at tree-level
by
the underlying theory, and those that are only generated by loops~\cite{jose}.
Under weak coupling conditions, tree-level operators are expected to be
dominant since the coefficients of
loop-generated operators have an additional $\sim (4\pi)^{-2}$
suppression factor. We therefore expect that observables affected by
tree-level-generated operators will be more sensitive to the heavy
physics. Using this and similar estimates we and
determine the reach in $M$ of any given experiment, and to select those
most sensitive to the new physics.

The paper is organized as follows: in the next section we briefly
describe the structure of an effective supersymmetric Lagrangian and
summarize the construction of supersymmetric gauge invariant operators.
In section \ref{one} we introduce the
supersymmetric model we use to exemplify the method, {\it{i.e.}}, the minimal
supersymmetric extension
of the Standard Model (MSSM) and the notation we use, and enumerate
all operators of dimensions 5 and 6.
In section \ref{two} we find those operators that are generated
at tree level from
vertices of a most general underlying supersymmetric theory.
As we mentioned before, these are expected to give heavy supersymmetric
corrections to processes with the MSSM as the low energy scenario. In section
\ref{three} we analyze a few examples which represent sensitive probes
of potential heavy contributions to low-energy physics.
Additional useful information is provided in the appendices.

%%%%%%%%%%%%%%%%%%%%%%%%%%%%%%%%%%%%%%%%%%

\section{Supersymmetric effective Lagrangians and supergauge invariant
operators} \label{zero}

%%%%%%%%%%%%%%%%%%%%%%%%%%%%%%%%%%%%%%%%%%

It is well known that in supersymmetric theories only the F component
($\th^2$ component) of chiral superfields
and the D component ($\tstbs$ component) of vector superfields are
invariant up to a total divergence under supersymmetry
transformations~\cite{wb}.
As will be explain in more detail in section \ref{two}, the effective
supersymmetric Lagrangian has the following structure
\begin{equation}
\La_{\rm {eff}} = \La_{\rm {MSSM}} + {1\over M} \left[ \sum_i a_i O_i^{(5)} +
\hbox{h.c.} \right] + {1\over M^2} \left[ \sum_i b_i O_i^{(6)} +
\hbox{h.c.} \right] + \cdots ,
\label{the.leff}
\end{equation}
where `h.c' denotes complex conjugation,
\{$O_i^{(5)}$\} and \{$O_i^{(6)}$\} are
gauge-invariant operators of dimension 5 and 6
that result after taking F or D components of suitable supersymmetric
operators, and \{$a_i$\}, \{$b_i$\}, {\it etc.}, are undetermined
(and, in general, complex) coefficients; the ellipsis denote higher
dimensional contributions.

All the operators are constructed out of the light superfields and
must be gauge invariant and respect supersymmetry. To illustrate the
procedure by means of which these operators are constructed consider two
chiral superfields
$\phi_{\bf n}$ and $\phi_{{\bf n}^\ast}$ transforming respectively
as the fundamental representation ${\bf n}$ of SU(N) and its complex conjugate
${\bf n^\ast}$, and the corresponding vector superfield $V$. Their
explicit transformation laws are
\begin{equation}
\phi_{\bf n} \rightarrow \phi'_{\bf n}=e^{-i\Lambda^a t^a}\phi_{\bf n}, \qquad
\phi_{{\bf n}^\ast} \rightarrow \phi'_{{\bf n}^\ast}=e^{-i\Lambda^a
(-t^{a\ast})}
\phi_{{\bf n}^\ast}, \qquad
e^V \rightarrow e^{V'}=e^{-i\Lambda^{a\ast} t^a} e^V e^{i\Lambda^a t^a}
, \end{equation}
where $t^a$ and $-t^{a\ast}$ are the (Hermitian) generators for ${\bf n}$ and
${\bf n^\ast}$ respectively; it follows that
\begin{equation}
\phi_{{\bf n}^\ast}^T \rightarrow {\phi^T_{{\bf n}^\ast}}'=\phi_{{\bf
n}^\ast}^T
e^{i\Lambda^a t^a} ,
\end{equation}
which will be useful below. Note that for the case of local gauge
transformations
the $\Lambda^a$ must be chiral superfields and therefore are necessarily
complex,
$\Lambda^a\ne\Lambda^{a\ast}$.

Using these definitions we can construct two types of invariants,
\begin{equation}
\phi_{{\bf n}^\ast}^T \phi_{\bf n} \qquad
\phi_{\bf n}^\da e^V \phi_{\bf n} .\label{inv.1}
\end{equation}
Since {\bf n} corresponds to the fundamental representation we also have
the additional invariant~\footnote{When the (super)fields transform
according to other irreducible representations additional invariants can
be constructed~\cite{weyl}.}
\begin{equation}
\e_{a_1 \cdots a_N }\phi^{a_1} \cdots \phi^{a_N } . \label{inv.2}
\end{equation}

All supersymmetric-invariant operators involving these fields
are constructed taking products of composites of the
form (\ref{inv.1}) and (\ref{inv.2}) and extracting their F or D
components.
In the MSSM both superfields associated with particles
and antiparticles are chiral; products of the form
$\phi_{{\bf n}^\ast}^T \phi_{\bf n}$ are then also chiral, and the
corresponding
F components will appear in the (effective) Lagrangian.
The terms of the form $\phi_{\bf n}^\da e^V \phi_{\bf n}$
always involve a chiral-antichiral product, hence its D component would be
chosen. A general operator will be a product of these gauge invariant
building blocks in which case each operator must be considered separately
in order to determine whether its
F or D component should be included.

Since both $ \phi_{\bf n}^\da \phi_{\bf n} $ and $ \phi_{{\bf n}^\ast}^T
\phi_{\bf n} $
are invariant under global gauge transformations (as is (\ref{inv.2})),
all operators can be obtained by first imposing global gauge
invariance and then replacing
\begin{equation}
\phi^\da \rightarrow \phi^\da e^{V_\phi}
\label{subs}
\end{equation}
for every chiral superfield, where
$V_\phi=\sum q^{(a)} V^{(a)} $ and the index
$a$ runs over the different gauge group factors present. The $q^{(a)}$
are the charges of $ \phi $ and
$V^{(a)}$ denote the corresponding Lie-algebra valued vector superfields.

It is important to recall at this point that, if an operator $O$ has
dimension $d$, then its F-component, $O_F$ has dimension $d+1$, while
its D-component, $O_D$ will have dimension $d+2$. For example, in order to
obtain the contributions to the effective Lagrangian of dimension 5 we need
chiral composites of dimension 4 and vector composites of dimension 3.
In the next section we construct all dimension 5 and 6 supersymmetric
contributions to the effective Lagrangian.

%%%%%%%%%%%%%%%%%%%%%%%%%%%%%%%%%%%%%%%%%%%%%%%%%%%%%%%%%%%%%%%%%%%%%%%%%%%%%

\section{MSSM and operators of dimensions 5 and 6}\label{one}

%%%%%%%%%%%%%%%%%%%%%%%%%%%%%%%%%%%%%%%%%%%%%%%%%%%%%%%%%%%%%%%%%%%%%%%%%%%%%

In this section we will construct the operators
of dimension $ \le6 $ appearing in the
effective Lagrangian. As mentioned in the introduction we assume that
the MSSM~\cite{mssm} describes the low energy
supersymmetric limit to lowest order in $M$ (the scale of the underlying
theory). This
model is invariant under local $SU(3)_C\times SU(2)_L\times U(1)_Y$;
the superfields involved are listed in table \ref{tab1}, these
include two chiral scalar superfields responsible for generating
masses for quarks and leptons.

\begin{table}
\[
\begin{array}{|c|c|c|c|c|c|r|}	\hline
\mbox{Superfield} & \mbox{spin-1} &\mbox{spin-1/2} & \mbox{spin-0}& SU(3) &
SU(2) & Y\\
\hline
Q & - & q_L=(u_L, d_L)^T & \qt_L & 3 & 2 & -1/6 \\
\U & - & \ub_L & \ut_L & 3^\ast & 1 & 2/3 \\
\D & - & \db_L & \dt_L & 3^\ast & 1 & -1/3 \\
L & - & l_L=(\nu, e)^T & \lt_L & 1 & 2 & 1/2\\
\E & - & \eb_L & \et_L & 1 & 1 & -1 \\ \hline
\Ho & - & \hot & \ho & 1 & 2 & 1/2 \\
\G & - & \gtt & \gt & 1 & 2 & -1/2\\ \hline
G & G_\mu & \lG & - & 8 & 1 & 0 \\
W & W_\mu & \lW & - & 1 & 3 & 0 \\
B & B_\mu & \lB & - & 1 & 1 & 0\\ \hline
\end{array}
\]
\caption{Assignment of representations and charges of MSSM superfields}
\label{tab1}
\end{table}

We now construct all operators involving the fields in
table \ref{tab1} and their covariant derivatives and whose F
and D components are of mass dimension 5 and 6.
To do this we follow the ideas outlined in the previous section and use
the canonical dimensions for superfields and derivatives as follows:
[chiral]$=1$, [vector]$=0$,
[gauge field strength]$=3/2$ and [covariant derivative]$=1/2$ (see, for
example, ~\cite{wb}).

Further conservation laws (such as those responsible for the suppression of
baryon number violating processes) are related
to additional gauge or discrete symmetries
\cite{wsy,parity} ({\it eg.} R-parity). These additional symmetries will be
imposed
 later. Anticipating these restrictions, however, we will ignore all operators
of
dimension 4 which violate R-parity (see appendix \ref{apc})
since they imply a short proton
lifetime~\cite{wsy,parity}. Additional
phenomenological constraints will also be considered at the end.

With these preliminaries we can now enumerate all dimension 5 and 6
operators of the MSSM. In constructing this list we have used the
equations of motion~\cite{arztone} to eliminate redundant contributions
(see appendix \ref{apa} for details). To simplify the expressions
it proves convenient to define
\begin{equation}
\begin{array}{ll}
V_Q =  G + W - {1 \over 6} B & \qquad V_L = W + { 1 \over 2} B \\
V_U = -G - { 2 \over 3} B & \qquad V_E = B \\
V_D = -G + { 1 \over 3} B & \qquad V_{H_{1, 2}} = \pm W + { 1 \over 2} B.
\end{array}
 \label{ves}
\end{equation}

In what follows $O_F$ indicates that the $F$ component of the operator
listed is to be included in (\ref{the.leff}); similarly $O_D$
indicates the $D$ component should appear in (\ref{the.leff}).
For compactness, we will refer to superfields $Q, \U, \D, L, \E$ as
``fermions'', $\Ho, \G$ as ``scalars'' and $G, W, B$ as ``vectors''.
For some of the operators listed below gauge invariance might allow more
than one index contraction; we choose to keep the indices implicit
for a simpler presentation.
In section \ref{three} we will calculate the contribution
of some of the operators to some physically interesting processes and we
will then make explicit the index structure of the relevant operators.

%%%%%%%%%%%%%%%%%%%%%%%%%%%%%%%%%%%%%%%%%%
\subsection{Operators of dimension 5}
%%%%%%%%%%%%%%%%%%%%%%%%%%%%%%%%%%%%%%%%%%

%%%%%%%%%%%%%%%%%%%%%%%%%%%%%%%%%
\subsubsection{Operators involving fermions only}
\begin{eqnarray}
\opf &=& QQ\U\D \label{Ffeone} \\
\opf &=& QQQL \label{propro}\\
\opf &=& \U\U\D\E \label{b.l.one}\\
\opf &=& Q\U L\E \label{Ffetwo}
\end{eqnarray}

%%%%%%%%%%%%%%%%%%%%%%%%%%%%%%%%%
\subsubsection{Operators involving fermions and scalars}

\begin{eqnarray}
\opf &=& QQQ\Ho \label{firstfescaone} \\
\opf &=& Q\U\E\Ho \label{secondfescaone}\\
\opf &=& LL\G\G \label{fescathree} \\
\opf &=& L\Ho\G\G \label{firstfescatwo}
\end{eqnarray}

%%%%%%%%%%%%%%%%%%%%%%%%%%%%%%%%%
\subsubsection{Operators involving scalars only}
\begin{eqnarray}
\opf &=& \Ho\Ho\G\G \label{scaone}
\end{eqnarray}

%%%%%%%%%%%%%%%%%%%%%%%%%%%%%%%%%
\subsubsection{Operators involving vectors only}

These operators must be constructed out of the
superfield strengths, which have mass dimension $3/2$
and we denote by $X_\al$. Lorentz
invariance allows only powers of $X^\al X_\al$ and its hermitian
conjugate. Only $[X^\al X_\al]_D$ are allowed, but these operators
correspond to total divergences.

%%%%%%%%%%%%%%%%%%%%%%%%%%%%%%%%%
\subsubsection{Operators involving fermions and vectors}

To obtain this set of operators we follow the procedure outlined in
section (\ref{zero}); after the replacement (\ref{subs}) we obtain
\begin{eqnarray}
\opd &=& QQ\De \label{feone}\\
\opd &=& Q\U\Le \\
\opd &=& \U\De\E \label{fetwo}
\end{eqnarray}
where $V_L$ and $V_D$ are defined in (\ref{ves}).

%%%%%%%%%%%%%%%%%%%%%%%%%%%%%%%%%
\subsubsection{Operators involving fermions, scalars and vectors}

Again following the procedure described in section \ref{zero} we obtain
\begin{eqnarray}
\opd &=& Q\U\He \label{fescaone} \\
\opd &=& Q\D\Ge \\
\opd &=& L\E\Ge \\
\opd &=& \E\Ho\Ge \\
\opd &=& \Ee\G\G \label{fescatwo}
\end{eqnarray}
where $V_{H_{1, 2}}$ and $V_E$ are defined in (\ref{ves}).

%%%%%%%%%%%%%%%%%%%%%%%%%%%%%%%%%
\subsubsection{Operators involving scalars and vectors}

No operators exist in this category since no gauge invariant expression
can be constructed out of three (MSSM) scalar superfields.

%%%%%%%%%%%%%%%%%%%%%%%%%%%%%%%%%%%%%%%%%%
\subsection{Operators of dimension 6}
%%%%%%%%%%%%%%%%%%%%%%%%%%%%%%%%%%%%%%%%%%

%%%%%%%%%%%%%%%%%%%%%%%%%%%%%%%%%
\subsubsection{Operators involving fermions only}

\begin{eqnarray}
\opf &=& QQQQ\U \label{bisFfeone}\\
\opf &=& QQ\U\U\E \\
\opf &=& \U\U\U\E\E \label{bisFfeR}\\
\opf &=& \D\D\D LL \label{bisFfetwo}
\end{eqnarray}

%%%%%%%%%%%%%%%%%%%%%%%%%%%%%%%%%
\subsubsection{Operators involving fermions and scalars}

\begin{eqnarray}
\opf &=& \U\D\D L \G \label{firstbisfescaone}\\
\opf &=& \D\D\D L\Ho \label{b.l.two}\\
\opf &=& Q\D LL \G \label{l.bminusl.one}\\
\opf &=& LLL\E\G \label{secbisfescaone}\\
\opf &=& \U\D\D\Ho\G \label{thirdbisfescaone}\\
\opf &=& \U\U\D\G\G \\
\opf &=& \D\D\D\Ho\Ho \\
\opf &=& Q\U L\G\G \\
\opf &=& Q\D L\Ho\G \label{fourbisfescaR} \\
\opf &=& LL\E\Ho\G \label{fourbisfescaone}\\
\opf &=& Q\U \Ho\G\G \label{fivebisfescaone} \\
\opf &=& Q\D\Ho\Ho\G \\
\opf &=& L\E\Ho\Ho\G \label{sixbisfescaone} \\
\opf &=& \E\Ho\Ho\Ho\G \label{sevenbisfescaone}
\end{eqnarray}

%%%%%%%%%%%%%%%%%%%%%%%%%%%%%%%%%
\subsubsection{Operators involving scalars only}

Chiral operators of this type must contain five chiral scalar
superfields; all such composites violate gauge invariance. There are no
operators in this category.

%%%%%%%%%%%%%%%%%%%%%%%%%%%%%%%%%
\subsubsection{Operators involving vectors only}

To obtain a dimension-6 contribution to the effective Lagrangian we need
an $O_F$ operator of dimension 5, or an $O_D$ operator of dimension 4. Neither
of these can be constructed purely out of powers of the field strength $
X_\al$.
There are no operators in this category.

%%%%%%%%%%%%%%%%%%%%%%%%%%%%%%%%%
\subsubsection{Operators involving fermions and vectors}

\begin{eqnarray}
\opd &=& \Qe\Qe QQ \label{bisfeone}\\
\opd &=& \Qe Q\Ue\U \\
\opd &=& \Qe Q\De\D \label{noseone}\\
\opd &=& \Ue \Ue\U\U \\
\opd &=& \Ue\U\De\D \\
\opd &=& \De\De\D\D \label{nosetwo}\\
\opd &=& QQ\Ue\Ee \label{b.l.three}\\
\opd &=& Q\Ue\De L \label{exc1}\\
\opd &=& \Qe Q\Le L \label{exc2}\\
\opd &=& \Qe Q\Ee \E \\
\opd &=& \Ue\U\Le L\\
\opd &=& \Ue\U\Ee \E \\
\opd &=& \De\D\Le L\\
\opd &=& \De\D\Ee \E\\
\opd &=& Q\D\Le\Ee \label{exc3}\\
\opd &=& \Le\Le LL \label{exc4}\\
\opd &=& \Le L\Ee \E \\
\opd &=& \Ee\Ee\E\E \label{bisfetwo} \\
\opd &=& \Qe\D\D L \\
\opd &=& \D\D\D\Ee \label{prodeone}\\
\opd &=& \Ue\D LL
\end{eqnarray}
where the various $V_I$ are defined in (\ref{ves}).

%%%%%%%%%%%%%%%%%%%%%%%%%%%%%%%%%
\subsubsection{Operators involving fermions, scalars and vectors}

\begin{eqnarray}
\opf &=& X_W^2 L H_2 \label{l.viol.one} \\
\opf &=& X_B^2 L H_2 \\
\opf &=& X_G^2 L H_2 \label{l.viol.two} \\
\opd &=& QQQ \Ge \label{bisfescaone} \\
\opd &=& \Qe\U\D\G \\
\opd &=& \Qe\D\D\Ho \\
\opd &=& \Qe Q L\G\\
\opd &=& \Ue\U L\G\\
\opd &=& \De\D L\G\\
\opd &=& Q\U\E\Ge\\
\opd &=& Q\D\Ee\G\\
\opd &=& \Ue\D L\Ho \\
\opd &=& \Le LL\G \\
\opd &=& Q\Ue\De\Ho \label{inc1}\\
\opd &=& Q\De\De\G \label{inc2}\\
\opd &=& \Qe Q L\He \label{inc3}\\
\opd &=& \Ue\U L\He \label{inc4}\\
\opd &=& \De\D L\He \label{inc5}\\
\opd &=& Q\D\Ee\He \label{inc6}\\
\opd &=& \U\De\Le\G \label{inc7}\\
\opd &=& \Le LL\He \label{inc8}\\
\opd &=& L\Ee\E\He \label{inc9}\\
\opd &=& L\Ee\E\G \label{fescathonebis} \\
\opd &=& \Qe Q \Ho\G \label{lost}\\
\opd &=& \Ue\U\Ho\G \\
\opd &=& \De\D\Ho\G \\
\opd &=& \Ue\D\Ho\Ho \\
\opd &=& \U\De\G\G \\
\opd &=& LL\He\G \\
\opd &=& \Le L\Ho\G \\
\opd &=& \Ee\E\He\Ho \\
\opd &=& \Qe Q \He\Ho \label{inc10}\\
\opd &=& \Qe Q \Ge\G \label{inc11}\\
\opd &=& \Ue\U\Ge\G \label{inc12}\\
\opd &=& \Ue\U\He\Ho \label{inc13}\\
\opd &=& \De\D\Ge\G \label{inc14}\\
\opd &=& \De\D\He\Ho \label{inc15}\\
\opd &=& \Ue\D\Ho\Ge \label{inc16}\\
\opd &=& LL\He\He \label{inc17}\\
\opd &=& \Le L\He\Ho \label{inc18}\\
\opd &=& \Le L\Ge\G \label{inc19}\\
\opd &=& \Ee\E\Ge\G \label{inc20}\\
\opd &=& \Ee\E\Ho\G \label{fescatwotwobis}\\
\opd &=& L\He\He\Ho \label{fescaoneth} \\
\opd &=& L\He\Ho\G \\
\opd &=& \Le\Ho\Ho\G \\
\opd &=& L\He\Ge\G \label{inc21}\\
\opd &=& L \Ge\G\G \label{bisfescatwo}
\end{eqnarray}
where $X_B$, $X_W$ and $X_G$ denote the field strengths associated with
the U(1), SU(2) and SU(3) gauge fields.

%%%%%%%%%%%%%%%%%%%%%%%%%%%%%%%%%
\subsubsection{Operators involving scalars and vectors}

\begin{eqnarray}
\opf &=& X_W^2 H_1 H_2 \\
\opf &=& X_B^2 H_1 H_2 \\
\opf &=& X_G^2 H_1 H_2 \\
\opd &=& \He\Ho\Ho\G \\
\opd &=& \Ho \Ge\G\G \\
\opd &=& \He\He\Ho\Ho \label{bisscaone} \\
\opd &=& \He\Ho \Ge\G \label{inc22}\\
\opd &=& \Ge\Ge\G\G \label{bisscatwo}
\end{eqnarray}
where $X_B$, $X_W$ and $X_G$ denote the field strengths associated with
the U(1), SU(2) and SU(3) gauge fields.

%%%%%%%%%%%%%%%%%%%%%%%%%%%%%%%%%%%%%%%%%%%%%%%%%%%%%%%%%%%%%%%%%%%%%%%%%%%%%

\section{Tree-level effective operators}\label{two}

%%%%%%%%%%%%%%%%%%%%%%%%%%%%%%%%%%%%%%%%%%%%%%%%%%%%%%%%%%%%%%%%%%%%%%%%%%%%%

When considering the phenomenological effects of the
above operators it must be emphasized that the contributions
of an operator will be
suppressed whenever it is generated by loops in the underlying
theory. Within our assumption of a weakly coupled underlying theory
the coefficient of any
loop--generated operator will be suppressed by a factor $ 1 / (16
\pi^2 ) $ together with a product of coupling constants. Operators
generated at tree level will only be suppressed by a product of
couplings. Because of this it becomes important to determine which of the
operators of the previous section are generated at tree level.

Following~\cite{aew} we first
enumerate all possible vertices present in a general renormalizable
supersymmetric
gauge theory. Using these vertices we construct all possible graphs
corresponding to effective contributions of dimensions 5 and 6. Since we
are interested in the effective theory at low energies we consider only
graphs with light external and heavy internal lines.

A general renormalizable superpotential contains terms of the form
$\phi^2$ and $\phi^3$ where $\phi$ denotes a fermion or scalar chiral
superfield.
Fermion-vector and scalar-vector interactions are generated by
$\phi^\da e^V \phi$, where $V$ denotes a vector superfield.
The expansion of the exponential will bring interactions of the form
$\phi^\da V^n \phi$ with $n$ an integer.
Finally, there are vector self-interactions coming from the kinetic
terms; the corresponding vertices contain three or more vector
lines.

Given this set of vertices we now discriminate between heavy and
light fields. We denote by $\psi_i, s_i, V_i$ with $i=l, h$
light (or heavy) fermion, scalar and vector superfields respectively.

A first restriction concerns effective operators involving vector superfields
only; they are generated by the kinetic terms and take the general form
(see for example~\cite{wb}),
\begin{equation}
f_{abc}(\bD^2 D^\al V_a) ( \bD^2V_b) (D_\al V_c), \;\;\;\;
f_{abe}f_{cde}(\bD^2 V_a)(D^\al V_b)\bD^2V_c(D_\al V_d)
\end{equation}
Since the light generators form a subalgebra any structure constant of
the form $f_{llh}$ must vanish. Therefore, as in the
non-supersymmetric case~\cite{aew}, the allowed vertices are of the form
\begin{equation}
V_lV_lV_l, \;\;\; V_lV_hV_h, \;\;\; V_hV_hV_h.
\end{equation}
For four-vector interactions only the following combinations of
structure constants are non-vanishing~\cite{aew}
$ f_{lll}f_{lll}, \ f_{hhl}f_{hhl}, \ f_{hhl}f_{lll}, \
f_{hhh}f_{hhh}, \ f_{lhh}f_{lhh}, \ f_{lhh}f_{hhh} $
leading to
\begin{equation}
V_lV_lV_lV_l, \;\;\; V_lV_lV_hV_h, \;\;\; V_lV_hV_hV_h, \;\;\; V_hV_hV_hV_h.
\end{equation}

A second restriction follows from the fact that the generators of the
light algebra cannot mix light with heavy chiral superfields. This disallows
vertices of the form $\psi_l^\da V_l^n \psi_h$ and $s_l^\da V_l^n s_h$
(n integer); similarly no vertex of the form $\psi^\da V_l^n s$ is present
in the theory, which
follows from the initial diagonalization of the kinetic terms.

The remaining restrictions apply to vertices
containing a single scalar superfield, which are generated by the
corresponding kinetic term when
a heavy scalar gets a vacuum expectation value
$v_h=\langle0|s_h|0\rangle$. Being interested in operators of dimension
$ \le 6 $ we need only consider diagrams with one heavy internal vector
line. This means that we can replace
\begin{equation}
e^{ V_l + V_h } \rightarrow \int_0^1 ds \; e^{ s V_l } V_h e^{(1-s) V_l }
\end{equation}
Using the fact that the
light generators annihilate $ v_h $ we have $ V_l v_h = 0 $. We also
note that given an broken generator $T_{\rm broken}$ the vector
$T_{\rm broken} v_h $ is the direction of a would-be Goldstone boson;
since these excitations transform among themselves under the unbroken
group we have $ s^\dagger T_{\rm broken} v_h = 0 $ and
$s^\dagger V_l T_{\rm broken} v_h = 0 $ for all physical
scalar superfields $s$. Collecting these results we get
\begin{eqnarray}
s_i e^{V_l + V_h } v_h
&=& s_i \int_0^1 ds \; e^{ s V_l } V_h e^{(1-s) V_l } v_h + O ( V_h^2 )
\nonumber \\
&=& O ( V_h^2 )
\end{eqnarray}
so that there are no vertices of the form $ s_i V_l^n V_h $.
There are no other general constraints to be derived from
gauge invariance.

Using the allowed vertices we now construct all possible diagrams
corresponding to contributions to the effective Lagrangian
of dimension 5 and 6 which are generated at tree level; the procedure
is straightforward, although lengthy. The diagrams are depicted in
figures \ref{fig:figthree} and \ref{fig:figfour}, the corresponding
operators are presented in table \ref{tab2}.

\setbox3=\vbox to 4 truein{\epsfysize=8.5 truein\epsfbox[-10 -432 602
360]{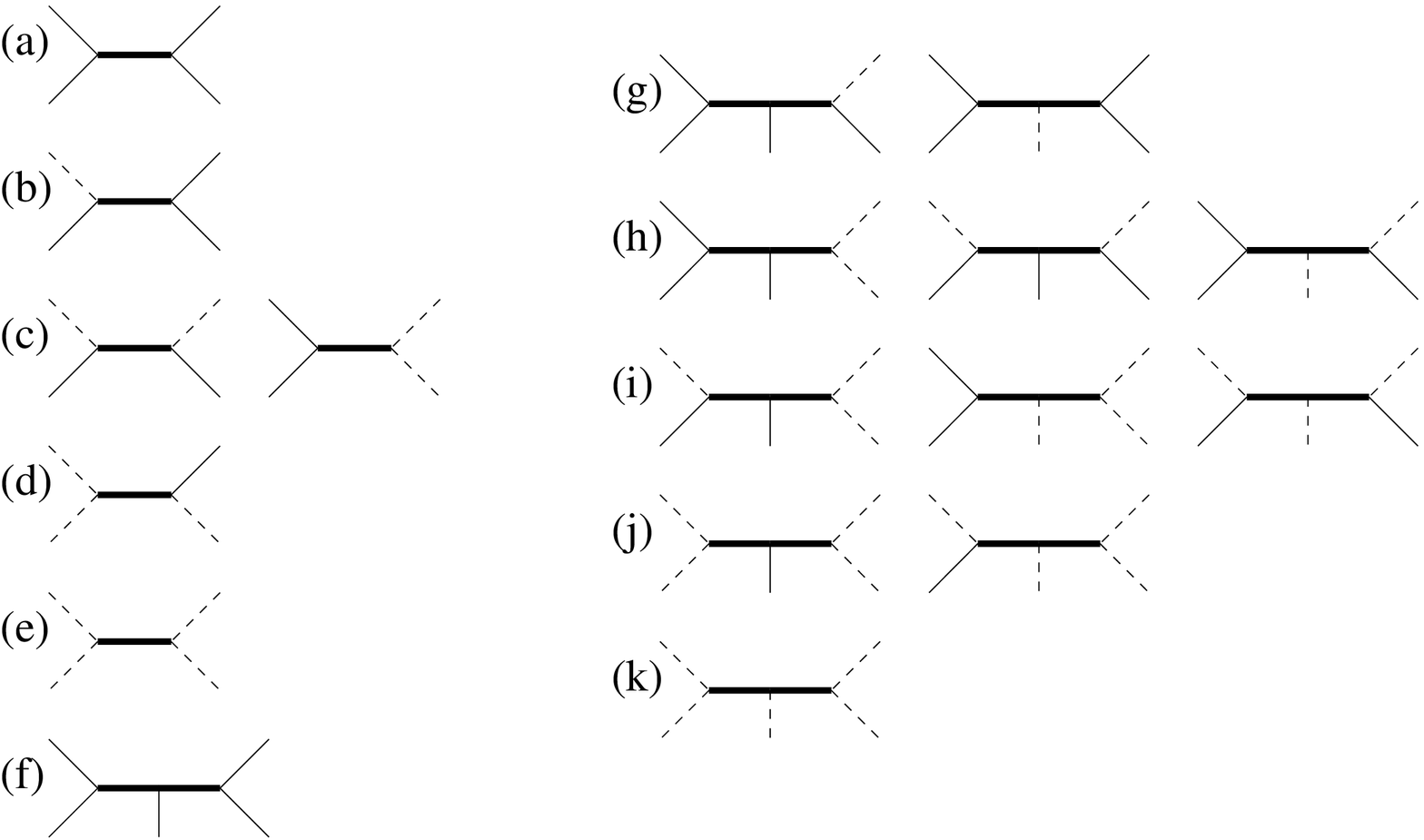}}
\begin{figure}[ht]
\centerline{\box3}
\caption{Diagrams corresponding to tree-level generated operators of type F.
Light fermions and scalars are represented by light full and dashed
lines respectively. heavy chiral fields (fermionic or scalar) are
represented by full heavy lines.
Diagrams (a)-(e) correspond to operators of dimension 5; (f)-(k) to
operators of dimension 6.}
\label{fig:figthree}
\end{figure}

\setbox3=\vbox to 3.3 truein{\epsfysize=7.2 truein\epsfbox[40 -812 652
-20]{pwff2.ps}}
\begin{figure}[ht]
\centerline{\box3}
\caption{Diagrams corresponding to tree-level generated operators of type D
(all of dimension 6). Same conventions for fermions and scalars as in
Fig.\ref{fig:figthree}.
Wavy lines with indices $i_l, \;l=1, 2$ on these diagrams
represent $i_l$ supervector lines converging to the same vertex.
Heavy chiral field propagators are denoted by heavy solid lines, heavy
vector propagators by heavy wavy lines.}
\label{fig:figfour}
\end{figure}

\begin{table}[ht]
\[
\begin{array}{|l|c|c|} \hline
\mbox{operator} & \mbox{diagrams} & \mbox{equation(s)}\\ \hline
d=5 \;\;\mbox{F-type} & & \\ \hline
\psi^4 &(\ref{fig:figthree}.a) &(\ref{Ffeone})-(\ref{Ffetwo}) \\
\psi^3 s &(\ref{fig:figthree}.b) &(\ref{firstfescaone})-(\ref{secondfescaone})
\\
\psi^2 s^2 &(\ref{fig:figthree}.c) &(\ref{fescathree}) \\
\psi s^3 &(\ref{fig:figthree}.d) &(\ref{firstfescatwo}) \\
s^4 &(\ref{fig:figthree}.e) &(\ref{scaone}) \\ \hline
d=6 \;\;\mbox{F-type} & & \\ \hline
\psi^5 &(\ref{fig:figthree}.f) &(\ref{bisFfeone})-(\ref{bisFfetwo})\\
\psi^4 s
&(\ref{fig:figthree}.g)&(\ref{firstbisfescaone})-(\ref{secbisfescaone}) \\
\psi^3 s^2
&(\ref{fig:figthree}.h)&(\ref{thirdbisfescaone})-(\ref{fourbisfescaone}) \\
\psi^2 s^3 &(\ref{fig:figthree}.i)
&(\ref{fivebisfescaone})-(\ref{sixbisfescaone}) \\
\psi s^4 &(\ref{fig:figthree}.j) & (\ref{sevenbisfescaone})\\
s^5 &(\ref{fig:figthree}.k) & - \\ \hline
d=6 \;\;\mbox{D-type} & & \\ \hline
\psi^4 e^V &(\ref{fig:figthree}.l) &(\ref{bisfeone})-(\ref{bisfetwo})\\
\psi^3 s e^V &(\ref{fig:figthree}.m) &(\ref{inc1})-(\ref{inc9})\\
\psi^2 s^2 e^V &(\ref{fig:figthree}.n)&(\ref{inc10})-(\ref{inc20}) \\
\psi s^3 e^V &(\ref{fig:figthree}.o) &(\ref{fescaoneth}), (\ref{inc21})\\
s^4 e^V &(\ref{fig:figthree}.p)&(\ref{bisscaone})-(\ref{bisscatwo})\\ \hline
\end{array}
\]
\caption{Tree-level generated operators; $ \psi , \ s , V $ denote,
respectively, fermion, scalar and vector superfields.} \label{tab2}
\end{table}

$O_D$ operators of dimension 5 are not generated at tree-level since
they involve
superfields with opposite chiralities, a type of vertex
that we have seen is not present in a supersymmetric potential.

In terms of components, operators (\ref{Ffeone}-\ref{scaone})
 contribute to the effective Lagrangian (\ref{the.leff})
terms with 2 scalars and 2 fermions and involve
either sfermions or higgsinos.
All these operators violate chiral symmetry and,
assuming naturality~\cite{tHooft}, are suppressed by powers of the fermion
masses.
Similarly (\ref{bisFfeone}-\ref{sevenbisfescaone}) generate terms with
three scalars and two fermions violating chiral symmetry and all
involve sfermions or higgsinos.

%%%%%%%%%%%%%%%%%%%%%%%%%%%%%%%%%%%%%%%%%%%%%%%%%%%%%%%%%%%%%%%%%%%%%%%%%%%%%

\section{Deviations from the MSSM}\label{three}

%%%%%%%%%%%%%%%%%%%%%%%%%%%%%%%%%%%%%%%%%%%%%%%%%%%%%%%%%%%%%%%%%%%%%%%%%%%%%

We have argued above that observables modified by tree-level-generated
operators are most sensitive to the heavy physics. We have also argued
that additional symmetries are required in order to prevent fast proton
decay. In this section we impose these further symmetries and determine
the observability of physics at the scale $M$ through the effects
of operators that are both R invariant and tree-level-generated. The
list of these operators is given in table \ref{tab3}.

\begin{table}[ht]
\[
\begin{array}{|l|c|c|} \hline
\mbox{operator} & \mbox{diagrams} & \mbox{equation(s)}\\ \hline
d=5 \;\;\mbox{F-type} & & \\ \hline
\psi^4 &(\ref{fig:figthree}.a) &(\ref{Ffeone})-(\ref{Ffetwo}) \\
\psi^2 s^2 &(\ref{fig:figthree}.c) &(\ref{fescathree}) \\
s^4 &(\ref{fig:figthree}.e) &(\ref{scaone}) \\ \hline
d=6 \;\;\mbox{F-type} & & \\ \hline
\psi^4 s
&(\ref{fig:figthree}.g)&(\ref{firstbisfescaone})-(\ref{secbisfescaone}) \\
\psi^2 s^3 &(\ref{fig:figthree}.i)
&(\ref{fivebisfescaone})-(\ref{sixbisfescaone}) \\ \hline
d=6 \;\;\mbox{D-type} & & \\ \hline
\psi^4 e^V &(\ref{fig:figthree}.l) &(\ref{bisfeone})-(\ref{bisfetwo})\\
\psi^2 s^2 e^V &(\ref{fig:figthree}.n)&(\ref{inc10})-(\ref{inc20}) \\
s^4 e^V &(\ref{fig:figthree}.p)&(\ref{bisscaone})-(\ref{bisscatwo})\\ \hline
\end{array}
\]
\caption{R-invariant tree-level generated operators; $ \psi , \ s, \ V $
denote, respectively, fermion, scalar and vector superfields.} \label{tab3}
\end{table}

The operators of type D in table \ref{tab3} contribute
three types of interactions (in terms of the component fields):
4-fermion interactions, terms with 2 fermions 2 scalars and one
covariant derivative, and terms with 4 scalars and two covariant
derivatives; none violate chiral symmetry. The operators
(\ref{inc10}-\ref{inc16}) and
(\ref{inc18}-\ref{inc20}) modify the couplings of the vector-boson to
the quarks and leptons and are correspondingly constrained by existing
data which implies$ M > 2\, $TeV
(taking the corresponding $b_i \sim 1 $ in (\ref{the.leff})
and assuming no cancelations)
\cite{wudka.grzadkowski}. Four-fermion interactions provide
similar bounds~\cite{SM,fourfermi}. Finally
(\ref{bisscaone})-(\ref{bisscatwo}) contribute two the $ \rho $
parameter which also provides a bound of the same order~\cite{SM}.
It is interesting to note that supersymmetry
and gauge invariance allows us to bound operators containing only
sparticles through current data.

We note that all these bounds merely state that $ M $ is greater than
(roughly) the scale at which low-energy supersymmetry is
expected to become manifest. Higher sensitivity can be achieved by
considering processes which are strictly forbidden with in the MSSM, but
which can be generated by the heavy dynamics. This possibility will be
considered in the next section.

\subsection{Rare processes}

The operators in table \ref{tab3} which respect the global symmetries of the
MSSM are subdominant compared to similar vertices induced by the MSSM
interactions themselves (this is the basic reason for above mild bounds on
$M$). There are a few operators, however, which violate some of the
global symmetries of the MSSM and the corresponding observables are much
more sensitive to $M$.

We will study three important examples: operators
contributing to proton decay, to flavor changing neutral currents (FCNC)
and to processes with lepton number violation.

\subsubsection{Proton decay}

Baryon and lepton number conservation are automatic in the Standard Model
after gauge invariance and renormalizability. However, this is not so
in supersymmetric models, where we need to invoke additional symmetries
\cite{wsy,parity} in order to forbid fast baryon decay through
dimension-four operators.
Discrete, global and gauge symmetries have been considered in the
literature, though only gauge symmetries are
respected by
gravitational interactions~\cite{wil}. Therefore only
the low-energy remnants of such extra gauge interactions appear
to be feasible candidates.
Here we use R-parity to forbid these dangerous operators~\cite{wsy,parity}.
After this is done, there are still operators of
dimensions 5 and 6 which violate baryon and lepton number
conservation (but conserve B-L), namely (\ref{propro}),
(\ref{b.l.one}), (\ref{b.l.three}) and
(\ref{exc1}); which involve only quark and lepton superfields
\begin{eqnarray}
d=5: & & \qquad [QQQL]_F, \;\;[\U\U\D\E]_F \\
d=6: & & \qquad [QQ\Ue\Ee]_D, \;\;[Q\Ue\De L]_D. \label{opi}
\end{eqnarray}
which are precisely the ones described in Refs.~\cite{wsy}. The $ d=5 $
operator $ [ Q Q Q H_1 ]_F $ violates baryon number but is not R-parity
invariant and is therefore not included in the above list. There are
various other operators, such as (\ref{feone}-\ref{fetwo}), which also
violate baryon number; these terms, however, are
loop generated and will give subdominant contributions.

The $ d = 5 $ operators are generated through the exchange of a
heavy chiral superfield,
whose mass we denote by $ M_S $, (diagrams (1.a)). The coefficient of
these operators will then be suppressed
by the corresponding Yukawa couplings~\cite{wsy}, denoted by \
$ y $; these operators have a prefactor $ \sim y^2 / M_S $. A specific
example is furnished by the $ SU(5) $ GUT~\cite{su5} where the heavy
chiral superfield corresponds to two scalar superfields transforming
according to the ${\bf5}$ and ${\bf\bar5}$ irreducible representations of $
SU(5)$.

 The $ d = 6 $ operators can be
generated through the vector superfield exchanges of diagrams
(1.l)~\footnote{The chiral exchanges in fig. (1.l) contribute only to
operators containing one or more vectors.}; their coefficient have the
form $ \sim g^2 / M_V^2 $ where $ M_V $ denotes the mass of the heavy
vectors and $g$ the corresponding gauge coupling constant. Within the $
SU(5) $ GUT~\cite{su5} such operators are generated by the exchange of
heavy vector superfields.

The baryon-violating vertices contained in the
d=5 effective operators involve
two sparticle external lines and therefore
induce proton decay only at loop-level (see for example~\cite{mssm,prode,wsy}
and references therein). For example, for the decay
$p\rightarrow K^+ \bar{\nu}$ the diagram
consists of one loop with two squarks and one gaugino propagator, and one
effective vertex of order $M^{-1}$. The corresponding amplitude is
$I_{(5)} \sim y^2 g^2 f/ ( 16 \pi^2 M_S ) $, where $g$ denotes
a gauge coupling constant and $f$ is a factor depending on the
light particles in the loop, explicitly~\cite{himuya}
$f\simeq m_{\tilde{g}}/m^2_{\qt}$ where $m_{\tilde{g}}$ and
$m_{\qt}$ are the gluino and squark masses respectively. For a rough
estimate we will take $I_{(5)}\sim y^2 g^2/ ( 16 \pi^2 \: M_S \: m_{susy} )$
where $m_{susy}$ denotes a typical scale of the light supersymmetry.

Regarding d=6 operators, only $[Q\Ue\De L]_D$ contributes to the decay
$p\rightarrow K^+ \bar{\nu}$.
Expanding this operator in terms of its components we obtain a term
\begin{equation}
[Q\Ud\Dd L]_D = \e_{ a b d } (u_a e - {d\, '}_a {\nu} )
\overline{\ub}_b \overline{s^c}_d
\end{equation}
where the Latin indices correspond to SU(3) and $s$ denotes the
s-quark field. The order of magnitude for the amplitude generated
by this vertex is
$I_{(6)}\sim g^2 / M_V^2$ where $ M_V $ denotes the vector mass and $g$
the corresponding gauge coupling constant (we will assume that all gauge
coupling constants are of the same order of magnitude).

The proton width is $ \sim \left| I_{(6)} + I_{(5)} \right|^2 m_p^5/ ( 8
\pi ) $, where $ m_p $ denotes the proton mass. The current lifetime
limit of $ 10^{32}$yr then implies $ M_V \gesim 2 \times 10^{15} \gev $ and $
M_S
\gesim 2 \times 10^{17} \gev $, having taken $ m_{susy} = 1 \tev $,
and using $ g = 0.656 $, $ y = 10^{-4} $~\cite{wsy}.

The
contribution from the d=5 operators dominate whenever $ M_V^2 /M_S > 2
\times 10^{13} \gev $. A specific
realization of this scenario occurs in $ SU(5)$ GUT models~\cite{su5} where
$M_S < M_V$ is assumed~\cite{wsy} (the estimates presented here may be
modified in specific models due to mixing-angle suppression or
cancelations). There also are scenarios in which
the contributions corresponding to $ d=5 $ operators are
suppressed such as in some $SO(10)$ grand unified
theories~\cite{baba}. In such cases the dominant effects
is generated by $d=6$ which occurs when $ M_V^2 /M_S < 2
\times 10^{13} \gev $; taking, for example, $M_S \sim 10^{18} \gev
$~\cite{baba} implies $ M_V \lesim 5 \times 10^{15} \gev $.

\subsubsection{Flavor changing neutral currents}

Flavor changing neutral currents are another important means to obtain
information regarding the physics underlying the MSSM~\cite{hake}.
As it is well known,
FCNC are suppressed in the Standard Model through the GIM
mechanism~\cite{GIM}.
In the MSSM, detailed analysis of the radiatively induced FCNC shows
that these are very small also (see~\cite{hake} and references therein).
In supersymmetric theories embedded in an underlying
GUT, large FCNC may be induced by off-generational Yukawa
couplings~\cite{bh}.

In what follows we consider the MSSM dominant contribution
to ${\rm K}^0-\bar{\rm K}^0$ mixing at the one-loop level to compare
with the tree-level contribution coming from effective vertices.
We compare these effects with the ones induced by
the effective operators
(\ref{bisfeone}), (\ref{noseone}) and (\ref{nosetwo})
\begin{equation}
\Qe\Qe QQ, \qquad \Qe Q\De\D, \qquad \De\De\D\D \label{opfcnc}
\end{equation}
which contribute through a 4-fermion contact interaction;
the operators (\ref{inc10}--\ref{inc16}) contribute through
a modification of the couplings of the vector-bosons to the quarks
(including a possible flavor-changing $Z$ coupling).
No dimension 5 or 6 F-type operator can contribute since these provide
2 or 3 sparticles, while we need
4 external fermion lines for this process.

The pure supersymmetric contribution generated by the MSSM
corresponds to the gluino-box diagram which has an amplitude of order
$I_{MSSM} \sim \xi^2/ ( 4\pi M_s)^2 $. The factors $(4\pi)^{-2}\sim1/160$
and $M_s^{-2}$ come from the loop integration and $\xi$ is an
additional suppression factor due to super-GIM mechanism
cancelations~\cite{hake} (which is also dependent on the squark mass
difference).
According to experiments, this amplitude is expected to be
$I_{MSSM}\le 10^{-15}$ GeV${}^{-2}$
(see for example~\cite{hake,gluglu}),
while the amplitude generated by the dimension 6 operators
is of order $I_{(6)} \sim M^{-2}$.
If we assume that there are no cancelations among contributions
from the three operators above, we obtain $M \ge 10^4$ TeV.

If flavor-changing heavy physics enters below this
threshold there must be GIM-like cancelations in the coefficients of
the dimension 6 operators contributing to FCNC. This added uncertainty
is always present when we consider operators which mix generations.

We note that all the four-Fermi effective operators in table
which give rise to FCNC are generated by the exchange of
chiral superfields. The flavor changing gauge vertices are produced
either by chiral or vector superfields

We briefly mention that the same approach can be used in lepton-flavor
violation. The relevant operators are (\ref{inc18}-\ref{inc20})
generated by heavy chiral or vector superfield exchanges. These
operators can induce $ \mu \rightarrow e \gamma $ with an amplitude $
\sim e v^2/ M^2 $ ($e$ denotes the electron charge), the corresponding
width is of order $ \sim \alpha m_\mu ( v/M)^4 $. The experimental
limits on the branching ratio are~\cite{SM} $ B ( \mu \rightarrow e
\gamma ) \le 4.9 \times 10^{-11} $, and correspond to $ M > 5.6 \times
10^8 \gev $. For a heavy mass $ \sim 10^{16} \gev $ the estimated
branching ratio lies 29 orders of magnitude below the experimental
limit. Note that these considerations do not apply to flavor violating
mass terms which can give measurable effects~\cite{bh.l}

\subsubsection{Lepton number violation}

Of the entries in table \ref{tab3} the
operators (\ref{fescathree}), (\ref{l.bminusl.one}),
(\ref{secbisfescaone}) and (\ref{inc17}) violate lepton number by two units
but conserve baryon number. Of these
\begin{equation}
\left[ LL\G\G \right]_F \label{l.viol}
\end{equation}
contains terms involving only known particles and generates a
Majorana mass for the neutrinos. This operator is generated through
the diagrams (1.c) corresponding to the exchange of a heavy scalar or
fermion superfield, the coefficient for this operator is then
proportional to a Yukawa coupling. The Majorana mass will then be
of order $ y v^2/M $, where $v$ denotes the vacuum
expectation value of the $ H_2 $ and $y$ a Yukawa coupling; $M$
corresponds to the mass of the heavy superfield~\footnote{The operator
(\ref{l.viol}) is R-parity invariant but violates other discrete
symmetries~\cite{parity} which can be used to eliminate it.}. There are
several other operators which violate lepton number such as
(\ref{l.viol.one}-\ref{l.viol.two}), these terms are loop generated (and
the corresponding contributions to the effective Lagrangian necessarily
involve sparticles) and
will give small contributions, assuming all lepton number violations
have the same characteristic scale.

Assuming no significant cancelations
among the various contributions to neutrinoless double-$\beta $ decay
current data implies~\cite{SM} $ y v^2/M \lesim 1 $ eV which corresponds
to $ M/y \gesim 6 \times 10^{13} \gev$. As a concrete example take the
underlying theory to be an $SO(10)$ GUT~\cite{so10} with the heavy
chiral field identified as the one containing the right-handed neutrino
$ \nu_R $. In this theory the coupling $y$ is related to the up-quark
Yukawa coupling so that $ y \sim 1, \ 0.006, \ 2 \times 10^{-5} $ for
the third, second and first generation respectively; correspondingly
we have $ M \gesim 6 \times 10^{13}, \ 4 \times 10^{11}, \ 10^9 \gev $.

%%%%%%%%%%%%%%%%%%%%%%%%%%%%%%%%%%%%%%%%%%%%%%%%%%%%%%%%%%%%%%%%%%%%%%%%%%%%%

\section{Conclusions}

%%%%%%%%%%%%%%%%%%%%%%%%%%%%%%%%%%%%%%%%%%%%%%%%%%%%%%%%%%%%%%%%%%%%%%%%%%%%%

In this paper an effective Lagrangian approach is presented which
allows the calculation of the
contributions to supersymmetric low-energy process due to a hypothesized
underlying supersymmetric heavy theory. This is a phenomenological
approach
in which a minimal amount of assumptions regarding the heavy physics
are made.

We have assumed that a low-energy theory such as the minimal supersymmetric
extension of the Standard Model (MSSM) is valid. Corrections
to this low-energy limit
are introduced by non-renormalizable operators constructed from MSSM
superfields.
These operators must have the symmetries of the low-energy theory, in this
case $SU(3)\times SU(2) \times U(1)$ and R-parity, the latter which
must be included to avoid fast nucleon decay.

For phenomenological reasons we divided the operators into those that
can be generated at tree level by the underlying physics and those which
must be generated via loops.
Under weak coupling conditions of the underlying
theory, tree-level operators are expected to be dominant
(for a given dimension) compared to loop-generated ones.
Having done this we can estimate the relevant contributions for a process, or
compare the coefficients of the operators and determine their relative
(estimated) strengths.

Existing data puts, for the most part, no stringent bounds on the scale
of heavy supersymmetry. We studied three exceptions: proton decay,
FCNC and lepton number violation. For the case of baryon number
violation the corresponding contributions have been extensively studied
in the literature~\cite{mssm,prode,hake,gluglu}; the above estimates
(validated by these detailed calculations) $ M_S \gesim 10^{17} \gev $
and $ M_V \gesim 10^{15} \gev $ for the masses of the heavy
scalars and vectors respectively. The constraints on FCNC imply $ M_S
\gesim 10^{13} \gev $. Finally neutrinoless double-$
\beta $ decay data requires $ M_S > 6 \times 10^{13} \gev $. In obtaining
these bounds we assumed that there are no significant cancelations
among the various contributions, when this is relaxed the bounds are
weakened.
Should the MSSM (or an alternative model) be discovered, the elucidation
of the various couplings and masses should considerably improve this
picture allowing for a variety of bounds obtained by considering
processes involving sfermions and other superparticles. The study of
rare processes, however, will provide the best bounds unless low-lying
mass thresholds exist.

The bounds derived constrain different types of heavy fields
depending on the diagrams responsible for the relevant operators.
It is important
to note that although the bounds obtained for $M$ using L violating operators
constrains the physics at which lepton number is violated, this does not
necessarily imply that {\em all} physics beyond low-energy
supersymmetry is
similarly bounded. Thus, while the bounds obtained
using rare processes
are quite severe, they do not necessarily
extend to the whole set of
operators obtained.

\section*{Acknowledgments}

J.W. is grateful to E. Ma for various discussions. This work is
supported in part through funds provided
by the U.S. Department of Energy under contract FDP-FG03-94ER40837.

%%%%%%%%%%%%%%%%%%%%%%%%%%%%%%%%%%%%%%%%%%%%%%%%%%%%%%%%%%%%%%%%%%%%%%%%%%%%%

\appendix

%%%%%%%%%%%%%%%%%%%%%%%%%%%%%%%%%%%%%%%%%%%%%%%%%%%%%%%%%%%%%%%%%%%%%%%%%%%%%

%%%%%%%%%%%%%%%%%%%%%%%%%%%%%%%%%%%%%%%%%%
\section{Equations of motion} \label{apa}
\setcounter{equation}{0}
\def\theequation{A.\arabic{equation}}
%%%%%%%%%%%%%%%%%%%%%%%%%%%%%%%%%%%%%%%%%%

When writing all possible effective operators from a given set of light fields
and covariant derivatives there may be redundant information; operators
which differ by terms vanishing when the equations of motion are used
are equivalent at the level of the S matrix (see for example
\cite{arztone} and references therein). One can then use the classical
equations of motion to reduce the actual set of operators to a
minimal set of inequivalent operators. In particular, for operators
that involve covariant derivatives, the equations of motion can be
used to reduce them to operators without covariant derivatives.
In what follows we sketch the way the substitution proceeds, without
trying to be exhaustive. We denote the supersymmetry-gauge
covariant derivative by $\na=e^{-V}D^\al e^V$ and $D^\al$ for the
covariant spinorial derivative.

Operators that involve
\begin{equation} \ns \p = e^{-V}D^2 e^V \p
\end{equation}
can be simplified by using the equations of motion.
The classical equation for $\p$ has the form
\begin{equation} e^V\ns \p= D^2 e^V \p = ({\pa \cal{W} \over \pa \p})^\da
\end{equation}
where $\cal{W}$ is the superpotential and involves operators up to dimension
three.

As an example we can consider operator (\ref{covder}) below. From the MSSM
superpotential (see appendix \ref{apc}), we obtain
\begin{equation} {\pa {\cal W} \over \pa Q} = a \U\G+b\D\Ho
\end{equation}
where $a, b$ are constants, and therefore
\begin{equation} (\Qd e^V\ns Q)^\da=a Q\U\G+b Q\D\Ho.
\end{equation}
The operators on the right are exactly (\ref{fourthree}) and
(\ref{fourfive}).
We also note that, in this particular case, the operators are
a total divergence after
taking the D-component on both sides, since the right-hand side is a
chiral operator.

Operators that involve the combination
\begin{equation} V_\al\na = {\bar{D}}^2\left[e^{-V}(D_\al e^V) \right] \na
\end{equation}
can be similarly simplified by using the equations of motion.

In what follows we enumerate all the operators involving covariant
derivatives and superfield strengths.
We use the notation $V_I^\al$ to represent the superfield
strength corresponding to superfields $V_I$ defined in (\ref{ves}).

%%%%%%%%%%%%%%%%%%%%%%%%%%%%%%%%%%%%%%%%%%
\subsection{$d=5$ operators}
%%%%%%%%%%%%%%%%%%%%%%%%%%%%%%%%%%%%%%%%%%

%%%%%%%%%%%%%%%%%%%%%%%%%%%%%%%%%
\subsubsection{Operators involving fermions, vectors and covariant derivatives}
\begin{eqnarray}
\opd &=& \Qe\ns Q \label{covder} \\
\opd &=& \Ue\ns \U\\
\opd &=& \De\ns \D\\
\opd &=& \Le\ns L \\
\opd &=& \Ee\ns \E
\end{eqnarray}
where the expression for the covariant derivative involves
those vector superfields that correspond to each case.\\

%%%%%%%%%%%%%%%%%%%%%%%%%%%%%%%%%
\subsubsection{Operators involving scalars, vectors and covariant derivatives}
\begin{eqnarray}
\opf &=& \Ho V_{\Ho}^\al\nas \G \\
\opd &=& \He\ns\Ho \label{scavecdone} \\
\opd &=& \Ge\ns\G \label{scavecdtwo}
\end{eqnarray}

%%%%%%%%%%%%%%%%%%%%%%%%%%%%%%%%%
\subsubsection{Operators involving fermions, scalars, vectors and covariant
derivatives}
\begin{eqnarray}
\opf &=& L V_L^\al\nas \G \\
\opd &=& \Le\ns \Ho
\end{eqnarray}
Similar comments as above for $V_L^\al$.

%%%%%%%%%%%%%%%%%%%%%%%%%%%%%%%%%%%%%%%%%%
\subsection{$d=6$ operators}
%%%%%%%%%%%%%%%%%%%%%%%%%%%%%%%%%%%%%%%%%%

%%%%%%%%%%%%%%%%%%%%%%%%%%%%%%%%%
\subsubsection{Operators involving fermions, vectors and covariant derivatives}
\begin{eqnarray}
\opf &=& Q\na\U Q\nas\D \\
\opf &=& Q\na\U L\nas \E \\
\opd &=& \Qd\nsb e^{V_Q}\ns Q \\
\opd &=& \Ud\nsb e^{V_U}\ns\U\\
\opd &=& \Dd\nsb e^{V_D}\ns\D \\
\opd &=& \Ld\nsb e^{V_L}\ns L \\
\opd &=& \Ed\nsb e^{V_E}\ns\E \\
\opd &=& \Qe V_Q^\al\nas Q \\
\opd &=& \Ue V_U^\al\nas\U \\
\opd &=& \De V_D^\al\nas\D \\
\opd &=& \Le V_L^\al\nas L \\
\opd &=& \Ee V_E^\al\nas \E
\end{eqnarray}

%%%%%%%%%%%%%%%%%%%%%%%%%%%%%%%%%
\subsubsection{Operators involving scalars, vectors and covariant derivatives}
\begin{eqnarray}
\opf &=& \Ho\na\Ho\G\nas\G \\
\opd &=& \Hd\nsb e^{V_{\Ho}}\ns\Ho \\
\opd &=& \Gd\nsb e^{V_{\G}}\ns\G \\
\opd &=& \He V_{\Ho}^\al \nas\Ho \\
\opd &=& \Ge V_{\G}^\al \nas\G
\end{eqnarray}

%%%%%%%%%%%%%%%%%%%%%%%%%%%%%%%%%
\subsubsection{Operators involving fermions, scalars, vectors and covariant
derivatives}
\begin{eqnarray}
\opf &=& Q\na\U E\nas\Ho \\
\opf &=& L\na L\G\nas\G \\
\opf &=& L\na\Ho\G\nas\G \\
\opd &=& \Ld\nsb e^{V_L}\ns \Ho \\
\opd &=& \Ld V_L^\al\nas \Ho
\end{eqnarray}

%%%%%%%%%%%%%%%%%%%%%%%%%%%%%%%%%%%%%%%%%%
\section{Dimension-four operators} \label{apc}
\setcounter{equation}{0}
\def\theequation{B.\arabic{equation}}
%%%%%%%%%%%%%%%%%%%%%%%%%%%%%%%%%%%%%%%%%%

We need dimension-four operators of the F-type to construct the
superpotential. The list is the following:
\begin{eqnarray}
\opf &=& Q\U\G \label{fourthree}\\
\opf &=& Q\D\Ho \label{fourfive}\\
\opf &=& L\E\Ho \label{fourfour}\\
\opf &=& Q\D L \label{fourone}\\
\opf &=& \U\D\D \\
\opf &=& LL\E \\
\opf &=& \Ho\Ho\E \label{fourtwo}
\end{eqnarray}
The first three operators conserve baryon and lepton number and enter
the MSSM Lagrangian.
Operators (\ref{fourone})-(\ref{fourtwo})
violate these symmetries and induce fast proton decay. They are excluded
by imposing and additional symmetry such as R parity (see section \ref{two})

%%%%%%%%%%%%%%%%%%%%%%%%%%%%%%%%%%%%%%%%%%%%%%%%%%%%%%%%%%%%

\end{document}